\documentclass[11pt,twoside]{article}
\usepackage{asp2010}

\resetcounters

\begin{document}

\title{The SDSS White Dwarf - M Star Library}
\author{Ren\'e Heller$^1$, Axel D. Schwope$^1$, and Roy H. {\O}stensen$^2$
\affil{$^1$Leibniz-Institut f\"ur Astrophysik Potsdam (AIP), An der Sternwarte 16, 14482 Potsdam, Germany}
\affil{$^2$Instituut voor Sterrenkunde, K.U. Leuven, Celestijnenlaan 200D, 3001 Leuven, Belgium}}

\begin{abstract}
The Sloan Digital Sky Survey (SDSS), originally targeted at quasi-stellar objects, has provided us with a wealth of astronomical byproducts through the last decade. Since then, the number of white dwarfs (WDs) with physically bound main-sequence star companions (mostly dM stars) has increased radically, allowing for fundamentally new insights into stellar physics. Different methods for the retrieval and follow-up analysis of SDSS WD-dM binaries have been applied in the literature, leading to a rising number of WD-dM catalogs. Here we present a detailed literature search, coupled with our own hunting for SDSS WD-dMs by color selection, the outcome being named the \textit{SDSS White Dwarf - M Star Library}. We also explain improvements of our automated spectral analysis method.
\end{abstract}

\section{Motivation}
\label{sec:introduction}

Among the compact binaries discussed in this conference, those composed of a white dwarf (WD) and a main-sequence (MS) companion, most of them being M dwarfs (dMs), rank among the least massive ones. Masses of free WDs cluster around $0.58\,M_\odot$ \citep{2007A&A...466..627H}, with $M_\odot$ as the mass of the Sun, but there are indications of higher WD masses in magnetic cataclysmic variables (mCVs) \citep{2000MNRAS.314..403R} in opposite to lower WD masses in post common envelope binaries (PCEBs) \citep{2011MNRAS.413.1121R}. The MS companions typically have masses between $0.1$ and $0.5\,M_\odot$ \citep[][H09 in the following]{2009A&A...496..191H}. With WDs being the common final state of MS star evolution \citep{2001PASP..113..409F} and dMs as the most abundant stars in the Milky Way, flanked by the fact that about half of WD progenitors exist in stellar binary systems \citep{1992ApJ...396..178F,2010ApJS..190....1R}, WD-dM binaries provide insights into standard paths of stellar evolution. Yet, these systems show a variety of phenomena, ranging from magnetic WD-dM interaction, over PCEBs and CVs to type Ia supernovae, mergers, accretion-induced collapses \citep{2002ApJ...565..430F}, and the emission of gravitational waves \citep{2010A&A...521A..85Y}, especially once both constituents in a close system have become compact objects. Moreover, both the WD and the dM once have formed from the same molecular cloud, thus virtually share same initial but completely different present compositions due to their dissimilar evolutionary stages. And since simulated WD evolutionary tracks offer the possibility to utilize the observed effective temperature and mass of the WD as a clock for the system, the age of widely separated MS companions, which would evolve similar to single stars, can be well constrained.

\begin{figure*}
  \centering
 \scalebox{0.395}{\includegraphics{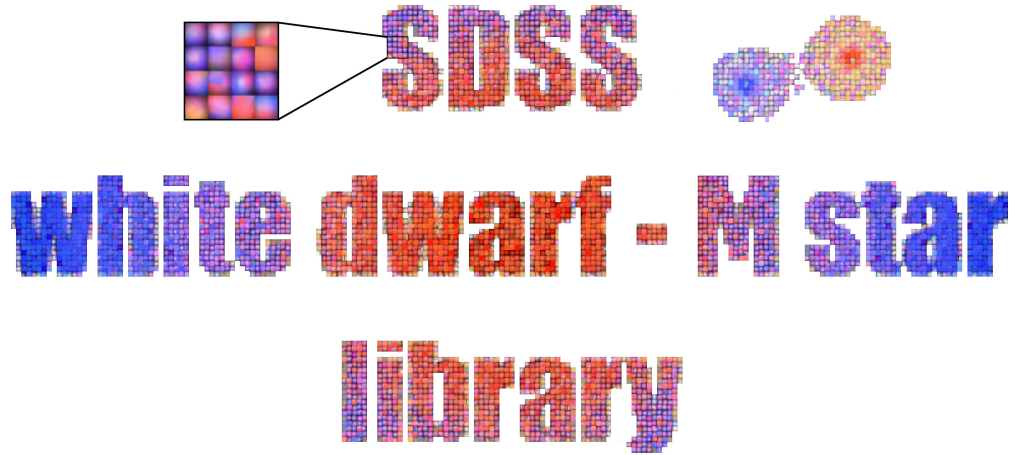}}
  \caption{Photometric data compilation of our 2537 WD-dM candidates. Systems with a dominant, red dM are found in the center, those governed by a blue-looking WD in the outer regions. Optically resolved pairs are preferably found in the transition region. A high-quality version is available at www.aip.de/People/RHeller.}
  \label{fig:library}
\end{figure*}

Since the advent of the Sloan Digital Sky Survey \citep[SDSS,][]{2000AJ....120.1579Y}, the number of known WD-dM pairs has increased dramatically. Since then, several studies have been dedicated to the identification and spectral analysis of these systems \citep{2003AJ....125.2621R,2004A&A...423..755P,2006A&A...454..617H,2007AJ....134..741S,2007MNRAS.382.1377R,2008MNRAS.388.1582L,2008A&A...486..843A,2009A&A...496..191H,2010A&A...520A..86Z,2010A&A...513L...7S,2010MNRAS.402..620R}. For our ongoing study, we compile these subsamples and use data of the recent SDSS Data Release 8 (DR8) \citep{2011ApJS..193...29A}. Here we present the outcome of the data acquisition and describe upgrades of the spectral analysis method with respect to H09.

\section{Catalog setup}

We first perform a query for WD-dM binary candidates in the DR8 using a color-search algorithm (H09). In the next step, we collect WD-dM candidates -- including PCEBs, CVs, and wide binaries -- from the publications mentioned in Sect.~\ref{sec:introduction} on systems from the SDSS and add the samples of \citet{2003ApJ...586.1356W}, \citet{2003A&A...406..305S}, \citet{2006MNRAS.370L..56N}, and \citet{2011ApJ...730...67B}, whose publications did not originally base on SDSS data. We then check this compilation for duplicates and reference the objects to further historical literature, such as \citet{1986AJ.....92..867G}, \citet{1999ApJS..121....1M}, \citet{1999yCat.3070....0L} \citet{2006ApJS..167...40E}, \citet{2004ApJ...607..426K}, \citet{2010yCat....102018R}, which mostly tabulate WDs. With this method, we accumulate 3540 WD-dM candidates, most of which were observed by the SDSS. This sample constitutes the \textit{White Dwarf - M Star Library}. We then use the DR8 query to search for sources at the objects' positions, which yields 2537 WD-dM candidates with SDSS spectra and photometric images (see Fig.~\ref{fig:library}), this sample being called the \textit{SDSS White Dwarf - M Star Library}. For 366 of these objects we find multiple spectra, thus more than tripling the repertoire of \citet{2007MNRAS.382.1377R}, forming a promising subsample for dynamical follow-up investigations, e.g. in search of PCEBs.

\begin{figure*}
  \centering
  \scalebox{0.35}{\includegraphics{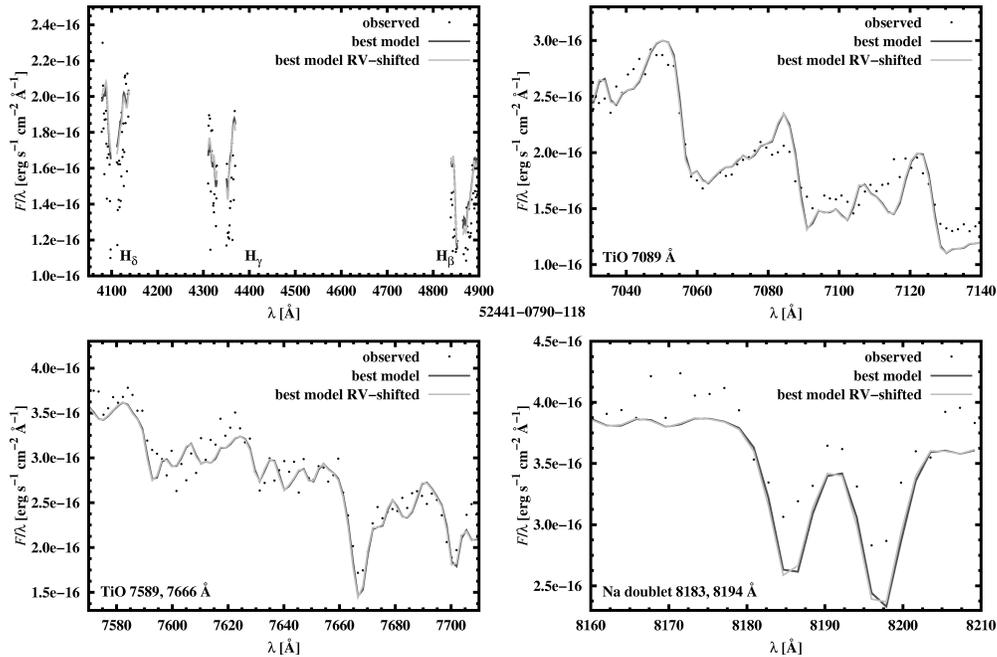}}
  \caption{RV fit for SDSSJ143746.68$+$573706.2 (52441-0790-118). The upper left panel shows the hydrogen lines used for the WD, the remaining panels show the spectral ranges and features used for the MS companion.}
  \label{fig:RV}
\end{figure*}

\section{Spectral analysis}

For our spectral analysis, we use a pre-computed grid of WD spectral models made available by D.~Koester \citep{1997ApJ...488..375F} as well as a grid of {\tt PHOENIX} models to fit the MS star \citep{jcam,1999ApJ...525..871H}. Our models are distributed in a five-dimensional parameter space, spanned by the effective temperatures of the two stars ($T_\mathrm{eff,WD}$ and $T_\mathrm{eff,dM}$), their surface gravities ($\log(g_\mathrm{WD})$ and $\log(g_\mathrm{dM})$), and the metallicity of the M dwarf ([Fe/H]$_\mathrm{dM}$). The mathematics of our $\chi^2$ fitting procedure is described in H09.

The main advancement of our parameter determination with respect to H09 is the derivation of the WD mass from the fitted $T_\mathrm{eff,WD}$ and $\log(g_\mathrm{WD})$ using evolution models from \citet{2010ApJ...717..183R}, rather than fixing the mass at $0.6\,M_\odot$. As a second improvement, we now fit for the radial velocities (RVs) of both objects simultaneously, using a $\chi^2$ minimization method (Fig.~\ref{fig:RV}). As a third enhancement, we measure the equivalent width of the H$\alpha$ emission line in order to assess the activity of the dM.

Analogously to the WD parametrization, we employ the fitted $T_\mathrm{eff,dM}$ and [Fe/H]$_\mathrm{dM}$ to estimate mass and radius of the MS companion from evolutionary models \citep{1997A&A...327.1039C}. Hence, we can estimate the distances of both binary constituents to Earth, ideally being equal for physically bound pairs.

\section{Results}

As a test we apply our {\tt python}-based Spectral Analyzing and Fitting Tool ({\tt SAFT}) to the eclipsing benchmark system SDSS\,\-J121258.25\-$-$012310.1 (SDSS1212$-$0123), whose parameters have been well constrained by \citet{2009A&A...495..561N} using detailed photometry and RV observations. A comparison between their parametrization of the system and ours is presented in Table~\ref{tab:SDSS1212}, with the results being in good agreement. The error bars from their targeted high-quality observations are much smaller than ours since we only decompose a single SDSS spectrum.

\renewcommand{\arraystretch}{1.1}
\begin{table}[t]
  \centering
   \caption{Comparison of the physical values for SDSS1212$-$0123 derived by \citet{2009A&A...495..561N} and in this study. $^1$ Conservative error estimates.}
   \label{tab:SDSS1212}

     \begin{tabular}{l|cl|cl}

    \hline

    \hline

    \hline \hline

    {\sc Parameter} & \multicolumn{2}{c|}{\sc Nebot-G\'omez-Mor\'an et al. (2009)} & \multicolumn{2}{c}{\sc This study$^1$} \\
    \hline

     $T_\mathrm{eff,WD}$                     & \hspace{0.6cm} $17\,700$ & $\pm~300$\,K         & \hspace{0.6cm} $17\,000$ & $\pm~1000$\,K       \\
     $M_\mathrm{WD}$                         & \hspace{0.6cm} $0.45$    & $\pm~0.01\,M_\odot$  & \hspace{0.6cm} $0.4$     & $\pm~0.1\,M_\odot$  \\
     $R_\mathrm{WD}$                         & \hspace{0.6cm} $0.017$   & $\pm~0.001\,R_\odot$ & \hspace{0.6cm} $0.02$    & $\pm~0.01\,R_\odot$ \\
     $\log(g_\mathrm{WD}/\mathrm{[cm/s^2]})$ & \hspace{0.6cm} $7.6$     & $\pm~0.1$            & \hspace{0.6cm} $7.5$     & $\pm~0.5$           \\
     $M_\mathrm{dM}$                         & \hspace{0.6cm} $0.275$   & $\pm~0.015\,M_\odot$ & \hspace{0.6cm} $0.15$    & $\pm~0.05\,M_\odot$ \\

    \hline
  \end{tabular}
\end{table}

\begin{figure*}[t]
  \centering
  \scalebox{0.44}{\includegraphics{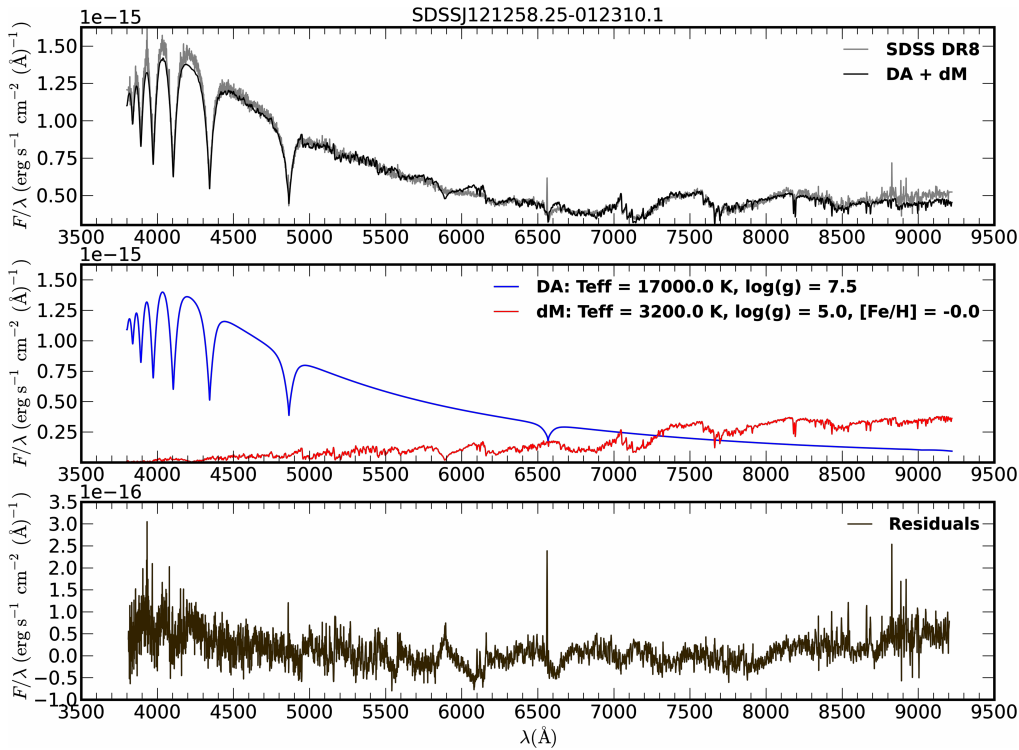}}
  \caption{Graphical output of {\tt SAFT} for the SDSS WD-dM binary SDSS1212$-$0123 (52367-0332-564). \textbf{Top:} Observed DR8 spectrum and combined WD-dM fit. \textbf{Center:} Best-fit decomposition of the spectrum with physical parameters in the legend. \textbf{Bottom:} Residuals.}
  \label{fig:SAFT}
\end{figure*}

In Fig.~\ref{fig:SAFT} we show the DR8 spectrum of SDSS1212$-$0123 with our fit as an overplot (top), as well as the spectral decomposition (center) and the residuals (bottom). With a reduced $\chi^2$ of 3.341 for 3832 wavelength-flux points in the SDSS file, corresponding to 1595 statistically independent data points (see Eq. (12) in H09), this fit is robust. In the residuals, the shortcomings of the {\tt PHOENIX} models can clearly be seen for $\lambda~\gtrsim~5500\,\AA$, as well as the H$\alpha$ emission line around $6564.6\,\AA$ (in vacuum), neglected by {\tt PHOENIX}.

During the library setup we encountered difficulties with the cross-matching of objects, partly caused by the changing SDSS name assignment for one and the same physical object among subsequent DRs. These variations probably stem from (\textit{i.}) proper motion of the objects, (\textit{ii.}) changing accuracy of the SDSS pointing \citep{2011ApJS..193...29A}, and (\textit{iii.}) occasional optical separation of the binary. As an arbitrary example, the object published by \citet{2003AJ....125.2621R} as SDSS\,J020806.39\-+001834.0, based on the Early DR, is termed SDSS\-\,J020806.3\-+001834.6 in DR1, SDSS\-\,J020806.31\-+001834.6 in DR5 \citep[but published as SDSS\-\,J020806.39\-+001834.0 by][]{2007AJ....134..741S}, SDSS\-\,J020806.31\-+001834.6 in DR7, and both as SDSS\-\,J020806.33\-+001834.5 and SDSS\-\,J020806.45\-+001833.7 in DR8. The latter double identification is probably due to the slight optical separation of the WD and the dM on the SDSS images, resulting in a multiple selection during the automatic SDSS targeting process.

\section{Conclusions and outlook}

Based on a comprehensive literature search, coupled with an independent search for WD-dM binaries in the color space, our \textit{White Dwarf - M Star Library} comprises the most complete library of WD-MS binaries up to date. In the following, we will apply an automated spectral analysis to the \textit{SDSS White Dwarf - M Star Library}, which will not only provide us with a standardized analysis of the catalog but will also enable us to sort out spurious objects, discern DA from DO and possibly other WD sub-types, and to further explore the RV and H$\alpha$ emission variability of the sub-sample with multiple spectroscopy. Therefore, it will be necessary to take into account the various selection effects of the publications we used to set up the library, i.e. to label CVs and consider optically resolved pairs. The latter objects serve as interesting targets for follow-up studies. Although their separation might induce a flux loss in the SDSS spectra of one component -- each of the 640 SDSS fibers covers a $1.5\,\arcsec$ radius on the celestial plane -- those systems which are resolved but with a separation well below $1.5\,\arcsec$ allow for an estimate of their minimum orbital period. Combined with the multi-epoch spectral analysis, triple systems could be identified.

The outcome of the spectral fitting to the whole \textit{SDSS White Dwarf - M Star Library}, in particular of the intrinsic and observational parameters derived, will be published in a subsequent paper.

In order to place a library as complete as possible at the disposal of the community, we encourage authors and astronomers to inform us about new WD-dM findings as well as about samples we might have been missing so far. Acknowledgements will be published.

\acknowledgements We thank A. Rebassa-Mansergas and his collaborators for sharing their sample prior to their original publication. R.~Heller is supported by the DFG grant SCHW~536/33-1. This research has made use of NASA's Astrophysics Data System Bibliographic Services and of the SIMBAD database, operated at CDS, Strasbourg, France. Funding for the SDSS and SDSS-II has been provided by the Alfred P. Sloan Foundation, the Participating Institutions, the National Science Foundation, the U.S. Department of Energy, the National Aeronautics and Space Administration, the Japanese Monbukagakusho, the Max Planck Society, and the Higher Education Funding Council for England. The SDSS Web Site is \url{http://www.sdss.org}.

\bibliography{RHeller-WDdM_SDSS}

\end{document}